\documentstyle[12pt,epsf]{article}

\newcommand{\f}{\frac}
\newcommand{\pdf}[2]{\frac{\partial#1}{\partial#2}}
\newcommand{\beq}{\begin{equation}}
\newcommand{\eeq}{\end{equation}}
\newcommand{\xit}{{\tilde \xi}}
\newcommand{\lambdat}{{\tilde \lambda}}
\newcommand{\Deltat}{{\tilde \Delta}}

\begin{document}

\begin{titlepage}
\vskip 1cm
\begin{flushright}
{\large
 UM-TH-95-10\\
 April 1995\\
 gr-qc/9504024\\}
\end{flushright}

\vskip 4cm

\begin{center}
{\Large\bf
 Interacting scalar fields in de~Sitter space}

\vskip 1cm

{\large
 Ganesh Devaraj\footnote{E-mail: {\tt gdevaraj@umich.edu}} and
 Martin B. Einhorn\footnote{E-mail: {\tt meinhorn@umich.edu}}}\\
\vskip 2pt
{\it Randall Laboratory of Physics, University of Michigan,\\
     Ann Arbor, MI 48109--1120, USA }
 \end{center}

\vskip .5cm

\begin{abstract}
We investigate the massless $\lambda \phi^4$ theory in de~Sitter
space.  It is unnatural to assume a minimally coupled
interacting scalar field, since $\xi=0$ is not a fixed point of the
renormalization group once interactions are included.
In fact, the only case where perturbation theory can be
trusted is when the field is non-minimally coupled at the minimum of
the effective potential.  Thus, in perturbation theory, there is no
infrared divergence associated with this scalar field.

\end{abstract}

\end{titlepage}
\setcounter{footnote}{0}
\setcounter{page}{2}
\setcounter{section}{0}
\setcounter{subsection}{0}
\setcounter{subsubsection}{0}

\section{Introduction}

Especially since the advent of the inflationary
cosmology,\cite{lindebook}
the study of de~Sitter space has become more relevant.  Inflation
aside, the well-known conundrum called the cosmological constant
problem,\cite{weinberg} (i.e., why the vacuum energy density is either
finely tuned or does not gravitate) is especially intransigent.
This unyielding theoretical puzzle  challenges our notions
of naturalness in field theory in other contexts, such as the
hierarchy problem in the Standard Model.   Various attempts have been
made to identify something that is ``wrong" with de~Sitter space
that would drive it toward Minkowski space.  In
particular, Ford and Parker \cite{Ford_Parker} pointed out that some
two-point functions for a massless, minimally coupled scalar (MMCS)
field are infrared divergent.
A closely related phenomenon is that the vacuum expectation value of
$\phi^2$ grows with
time \cite{Linde}\cite{Starobinsky}\cite{Vilenkin_Ford},
\begin{equation}
\langle \phi^2 \rangle = \frac{H^3}{4 \pi^2} t .
\label{eq:phisq}
\end{equation}
Here we have parameterized the de~Sitter metric as
\beq
ds^2=dt^2-e^{2Ht}d{\bf x}^2.
\eeq
The result, Eq.(\ref{eq:phisq}), is surprising in that it is {\it
not} de~Sitter invariant.
Because of the infrared divergence, $\langle \phi^2 \rangle$ is
calculated either by introducing a physically well motivated
infrared cutoff \cite{Linde} or by choosing
other physically acceptable vacuum states which are free of infrared
divergences \cite{Ford_Parker}.  Either way the result is
Eq.(\ref{eq:phisq}) with the possible addition of a constant (whose
effect is to shift the origin of the time coordinate) and of terms
which decrease exponentially in time \cite{Ford}.

Ford \cite{Ford} has used this growth in time of
$\langle\phi^2\rangle$ to argue that an interacting theory which
includes a massless, minimally coupled scalar field may, under certain
circumstances, cause a decrease in the cosmological constant
from a large value to a small value, thus making de~Sitter space
unstable.  We note that this
problem with the infrared divergence does not arise for a
non-minimally coupled scalar field or for a massive scalar field.
In these cases the de~Sitter invariant vacuum is free of infrared
problems, and $\langle \phi^2 \rangle$ does not grow with time.  This
property is crucial to the conclusion of this paper.

In this paper we discuss the one-loop effective potential for
massless $\lambda \phi^4$ theory in de~Sitter space and
show that, in all cases where perturbation theory is valid,
the field $\phi$ is not minimally coupled, that is, the coupling to
the curvature, $\xi$, is not equal to zero,
at the minimum of the effective potential.
Although we only treat the $\lambda\phi^4$ theory in
detail, the lessons learnt here are applicable to other models.
The upshot of our analysis
is this -- once the field $\phi$ becomes an interacting field, it
cannot be considered minimally coupled
and, hence, Eq.(\ref{eq:phisq}) does not hold for
an interacting scalar field in de~Sitter space.  Thus in such a
situation, there is no infrared problem.  We have not
considered the feedback on the metric, which may lead to other
instabilities \cite{Mottola}.

\section{The Effective Potential}

As the simplest model, consider
a massless, non-minimally coupled, $\lambda \phi^4$ theory in
de~Sitter space described by the Lagrangian density,
\beq
L=\sqrt{-g}\left[\frac{1}{2}g^{\mu\nu}\partial_\mu\phi \partial_\nu\phi
-\frac{\xi R}{2}\phi^2 - \frac{\lambda}{4!}\phi^4 + \rm{c.t.}\right]
\label{eq:L}
\eeq
In the zeta-function scheme, the one-loop effective potential is
\cite{Allen},
\beq
V_{eff}(\phi)=\frac{\xi R}{2}\phi^2+\frac{\lambda}{4!}\phi^4
-\frac{1}{2\Omega}\left[\zeta'(0)+\zeta(0)
\ln\left(\f{\mu^2}{H^2}\right)\right]
\label{eq:Veff}
\eeq
where $\Omega=\f{8}{3} \pi^2 \f{1}{H^4}$, $R=12 H^2$,
$$
\zeta(0) = \frac{\Delta^2}{12}-\f{\Delta}{24}-\f{17}{2880},
$$
\begin{eqnarray}
\zeta'(0) &=& -\f{1}{3}\biggl[\int_\f{1}{2}^{\f{1}{2}+\sqrt\Delta}
U(U-\f{1}{2})(U-1)\Psi(U) dU \nonumber \\
&&\mbox{}+ \int_\f{1}{2}^{\f{1}{2}-\sqrt\Delta}
U(U-\f{1}{2})(U-1)\Psi(U) dU \biggr]+\f{\Delta^2}{12}+\f{\Delta}{72},
\nonumber
\end{eqnarray}
(up to an additive constant) and,
\beq
\Delta\equiv \f{9}{4}-\f{V''_{tree}}{H^2}=
\f{9}{4}-\f{1}{H^2}\left(\xi R+\f{\lambda}{2}\phi^2\right)
=\f{9}{4}-12\xi-\f{\lambda}{2}\f{\phi^2}{H^2}.
\label{eq:Delta}
\eeq
Here $\mu$ is the (arbitrary) renormalization scale, on which
the renormalized coupling constants $\xi(\mu)$, and
$\lambda(\mu)$ depend.  Depending on the value of
$\Delta$, the integration is along the line $Re(U)=1/2$ (when $\Delta$
is negative) or along the real axis (when $\Delta$ is positive).  The
digamma function (defined by $\Psi(U)=d\Gamma(U)/dU$) has simple poles at
$U=0,-1,-2,..$.  The pole at $U=0$ is cancelled in the integrand.
The integration around the poles at $U=-1,-2,...$ is performed by
deforming the contour under the poles in the complex $U$ plane.
The imaginary part of $V_{eff}$ is given by
$i \pi \times$ the residue of the poles \footnote{For the
interpretation of ${\rm Im}\{V_{eff}\}$, see Weinberg and Wu
\cite{W&W}.}.

Now we want to investigate where the minimum of $V_{eff}$
occurs and whether it is consistent to have $\xi=0$ at the global
minimum of $V_{eff}$. In the models
that Ford \cite{Ford} considered, the renormalized value of $\xi$
was set to zero at some renormalization scale $\mu$,
and Eq.(\ref{eq:phisq}) was
used for $\langle \phi^2 \rangle$.  But since the field $\phi$ is an
interacting field in the models considered in Ref.~\cite{Ford}, the
beta function for $\xi$ is non-zero, and more importantly, $\xi=0$ is
not a fixed point of the renormalization group. Therefore
it is important to verify whether one can arrange for the renormalized
value of $\xi$ to be zero at the global minimum of the effective
potential (the vacuum).  What we find is that perturbation theory
breaksdown in theories where $\xi$ vanishes at the minimum.  Also,
normalizing the theory at the scale where $\xi$ vanishes, also leads
to the breakdown of perturbation theory.  We will illustrate in the
simplest case of $\lambda\phi^4$, but we expect that this model is
prototypical of more general interacting scalar theories, and the
results we obtain will be directly relevant for such theories (the
interacting scalar model in Ref.~\cite{Ford} for example).

Now let us examine the effective potential in the limits $\phi^2\gg H^2$
and $\phi^2\ll H^2$ to see what choices we have to make in these regions
for $\mu^2$ so as to obtain a satisfactory perturbation expansion.  In
the limit $\phi^2\gg H^2$, the asymptotic behaviour of the digamma
functions is logarithmic.  In this limit $\zeta'(0)$ tends to,
\beq
\zeta'(0) \rightarrow -\zeta(0)
\ln\left(\f{\lambda}{2}\f{\phi^2}{H^2}\right)
\eeq
Thus the one loop correction in this limit tends to \cite{Allen},
\beq
\rightarrow \f{1}{2 \Omega} \zeta(0)
\ln\left(\f{\lambda}{2}\f{\phi^2}{\mu^2}\right)
\eeq
which is of course the one loop correction in flat space.  To keep the
log under control in this region we should choose $\mu^2 \approx
\phi^2$.

In the region $\phi^2\ll H^2$, the behaviour of the digamma functions is
not logarithmic.  To examine the behaviour of $V_{eff}$ more precisely
in this region, it is simpler, and for our purposes sufficient to look
at $dV_{eff}/d\phi\equiv V'_{eff}$.  Differentiating
Eq.(\ref{eq:Veff}) we obtain,
\begin{eqnarray}
V'_{eff}&=&\phi\biggl[\xi R + \f{\lambda}{6}\phi^2 -
\f{\lambda R}{384 \pi^2} \biggl[\left(\Delta-\f{1}{4}\right)
\biggl\{\Psi\left(\f{1}{2}+\sqrt{\Delta}\right) \nonumber \\
&&~~~~~~~~~~~~~~~+\Psi\left(\f{1}{2}-\sqrt{\Delta}\right)
-1- \ln\left(\f{\mu^2}{H^2}\right)
\biggr\} -\f{1}{3} \biggr] \biggr]
\label{eq:Vprime_nonimp}
\end{eqnarray}
For $\phi^2/H^2\rightarrow 0$, $\Delta\rightarrow 9/4-12 \xi$.  So the
behaviour of the digamma functions is regular and they tend to a
constant value as $\phi^2/H^2\rightarrow 0$ (except for particular
values of $\xi$ which we will discuss later).  Therefore, for
$\phi^2/H^2 \ll 1$, if we choose $\mu^2=H^2$, there will be no large
log in the expression.  Hence, the Hubble scale, $H$, acts as an
infrared cutoff.  Upto small factors, this choice for the infrared
cutoff matches with that of Ref.\cite{Linde2} and
Ref.\cite{Taylor_Veneziano}.

Therefore the perturbation expansion will be improved if we exploit
the renormalization group to resum the large logs for $\phi^2>H^2$.
For all $\phi^2\le H^2$ we may simply choose to normalize at
$\mu^2=H^2$.

\section{The Renormalization Group Equation}

The renormalization group equation satisfied by $V_{eff}$
is,\cite{Elizalde_Odintsov}
\beq
\left(\mu\f{\partial~}{\partial\mu}+\beta_\xi \pdf{~}{\xi} +
\beta_\lambda \pdf{~}{\lambda} - \gamma\phi \pdf{~}{\phi} \right)
V_{eff}(\phi,\xi,\lambda,\mu) = 0
\label{eq:RGE}
\eeq
The beta functions calculated to one-loop using the background field
method are \cite{Elizalde_Odintsov}\cite{bos},
\beq
\beta_\xi=\f{\lambda}{16
\pi^2}\left(\xi-\f{1}{6}\right)
{}~~~~~~~~~~~\beta_\lambda=\f{3\lambda^2}{16 \pi^2}
{}~~~~~~~~~~~\gamma=0~~~~
\eeq

Note that $\xi=0$ is not a fixed point, and the conformally coupled
value, $\xi=1/6$, is a fixed point at one loop.
What we are primarily interested in is the position of
the minimum of $V_{eff}$.  Instead of applying the RG
to $V_{eff}(\phi)$ and then differentiating, it is simpler to apply the
RG to $V'_{eff}(\phi)$ and work with the
RG-improved $V'_{eff}$. Differentiating (\ref{eq:RGE}) with respect
to $\phi$, and using the fact that $\gamma=0$ to one-loop, we obtain the
RGE satisfied by
$V'_{eff}$.
\beq
\left(\mu\f{\partial~}{\partial\mu}+\beta_\xi \pdf{~}{\xi} +
\beta_\lambda \pdf{~}{\lambda}\right)
V'_{eff}(\phi,\xi,\lambda,\mu) = 0
\eeq
A solution to this equation is,
\beq
V'_{eff}(\phi,\xi,\lambda,\mu)=V'_{eff}(\phi,\xi(\phi),\lambda(\phi),\phi)
\eeq
where the running couplings are given by,
\begin{eqnarray}
\xi(\phi)&=&\f{1}{6}+\f{\xi(\mu)-\f{1}{6}}{(1-\f{3\lambda(\mu) t}{16
\pi^2})^\f{1}{3}} \label{eq:runningxi} \\
\lambda(\phi)&=&\f{\lambda(\mu)}{1-\f{3\lambda(\mu) t}{16 \pi^2}}
\label{eq:runninglambda} \end{eqnarray}
where $t=\f{1}{2}\ln(\phi^2/\mu^2)$, $\xi(\mu)\equiv\xi$, and
$\lambda(\mu)\equiv\lambda$.
Thus the RG improved $V'_{eff}$ is,
\begin{eqnarray}
V'_{eff}&=&\phi\biggl[\xi(\phi) R + \f{\lambda(\phi)}{6}\phi^2 -
\f{\lambda(\phi) R}{384 \pi^2} \biggl[\left(\Delta(\phi)-\f{1}{4}\right)
\biggl\{\Psi\left(\f{1}{2}+\sqrt{\Delta(\phi)}\right) \nonumber \\
&&~~~~~~~~~~~~~~~+\Psi\left(\f{1}{2}-\sqrt{\Delta(\phi)}\right)
-1- \ln\left(\f{\phi^2}{H^2}\right)
\biggr\} -\f{1}{3} \biggr] \biggr]
\label{eq:Vprime}
\end{eqnarray}
By $\Delta(\phi)$ we mean, use Eq.(\ref{eq:Delta}) with the
running coupling constants.  Since $\gamma=0$, the full content of the
RGE is obtained by setting $\mu=\phi$ in Eq.(\ref{eq:Vprime_nonimp}).

The solution we have obtained, Eq.(\ref{eq:Vprime}), is the solution
to the RG equation with beta functions calculated using a mass
independent renormalization prescription.  Hence, physical ``mass
thresholds'' are not automaticaly incorporated and so they have to be
separately incorporated.  Drawing upon the analysis from the previous
section, we see
that Eq.(\ref{eq:Vprime}) should be used only for scales $\phi^2\ge H^2$.
For scales $\phi^2<H^2$, we simply choose $\mu^2=H^2$.  So
the explicit log term goes to zero and the couplings remain frozen at
the values $\xi(H)$ and $\lambda(H)$ for all $\phi^2<H^2$.  Therefore
for $\phi^2\ge H^2$, $V'_{eff}$ is given by Eq.(\ref{eq:Vprime}), and
for $\phi^2<H^2$, $V'_{eff}$ is given by,
\begin{eqnarray}
V'_{eff}&=&\phi\biggl[\xi(H) R + \f{\lambda(H)}{6}\phi^2 -
\f{\lambda(H) R}{384 \pi^2} \biggl[\left(\Delta(H)-\f{1}{4}\right)
\biggl\{\Psi\left(\f{1}{2}+\sqrt{\Delta(H)}\right) \nonumber \\
&&~~~~~~~~~~~~~~~~~~~~~~~~~+\Psi\left(\f{1}{2}-\sqrt{\Delta(H)}\right)
-1\biggr\} -\f{1}{3} \biggr] \biggr]
\label{eq:Vprime_H}
\end{eqnarray}
Here, by $\Delta(H)$ we mean, use Eq.(\ref{eq:Delta}) with the
couplings normalized at $H$.  Henceforth by $V'_{eff}$, we will mean
$V'_{eff}$ given by Eq.(\ref{eq:Vprime}) and Eq.(\ref{eq:Vprime_H}),
depending on the value of $\phi^2$.  The RG improved $V_{eff}$ can be
obtained by integrating $V'_{eff}$.

\section{The Form of the Effective Potential}

The form of $V_{eff}$ is determined by the parameters :
$\xi(\mu)$, $\lambda(\mu)$, and $\mu/H$, or equivalently by $\xi(H)$
and $\lambda(H)$.
Depending on the values of these dimensionless coupling constants,
$V_{eff}$ takes on one of the two forms labeled (a) and (b)
in Figure 1 (We are ofcourse only dealing with theories in which
$\lambda(H)/(16\pi^2)\ll 1$).
\begin{figure}
\centering
\epsfxsize=5in
\hspace*{0in}
\epsffile{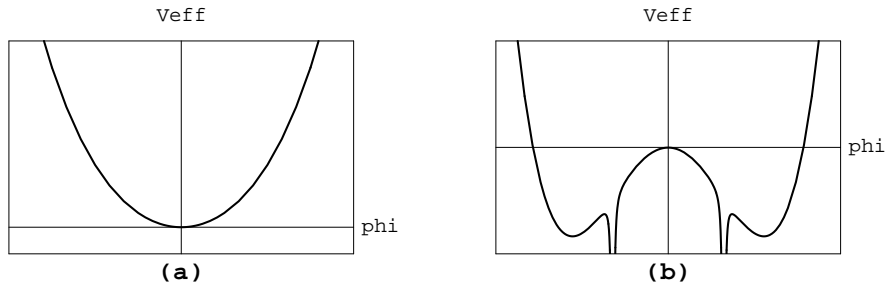}
\caption{Plot of $V_{eff}$ verses $\phi$ for the two perturbatively
valid cases.}
\label{fig1}
\end{figure}
Note that we have subtracted the appropriate constant so that
$V_{eff}$ is zero at $\phi=0$.  Figure 1(a) shows the symmetric case,
and Figure 1(b) shows the spontaneously broken case, that is, when the
symmetry is spontaneously broken.
It turns out that there is another possibility.
For some values of $\xi(H)$, and $\lambda(H)$, perturbation theory
breaks down in the neighborhood of the
`apparent minimum' of $V_{eff}$.  Therefore in this case we can not
identify the minimum of the effective potential using a perturbation
calculation.

Before we go further we will describe qualitatively
how this breakdown of perturbation theory shows up
in the one-loop effective potential.
Consider the tree potential with the running couplings (recall that we
keep the couplings frozen for $\phi^2<H^2$).  We will refer to this
potential as $V_0$ henceforth.  Suppose $V_0$ has a
minimum at $\phi=\phi_0$, which could be zero or non-zero.
We would expect that as long as $\lambda(\phi_0)/(16 \pi^2) \ll 1$,
perturbation theory could be trusted, and $V_0$ would only be
modified slightly upon the inclusion of the full one loop
corrections.\footnote{There is no
possibility for a Coleman-Weinberg \cite{Coleman_Weinberg} type of
situation here.}  But this is not always true.
For some range of values of the couplings, one of the poles in
$V'_{eff}$ (the pole that occurs when the curvature
of the tree potential, $V''_{tree}(\phi)$, vanishes)
significantly affects the structure of the effective
potential near this apparent minimum, $\phi_0$.  This implies a
breakdown of perturbation theory in the neighborhood of $\phi_0$,
even though $\lambda$ may be perturbatively small.
We note that a divergence at $V''_{tree}(\phi)=0$ also occurs in the
flat space effective potential, but only at higher orders
\cite{FJSE}.  As we mentioned, this
situation can arise when the minimum of $V_0$, $\phi_0$, is at the
origin or away from the origin.  We will discuss both these cases
below.

We will treat the symmetric case first.  We will show that
a necessary condition for the symmetric case is that $\xi(H)>0$.  But
$\xi(H)$ can not be too small (we will make this
quantitative later), since, if the curvature of the effective
potential becomes
too flat, the fluctuations become uncontrolled leading to a
breakdown in perturbation theory.  This is the physical reason for the
appearance of the singularity at $V''_{tree}(\phi)=0$ in the
perturbatively calculated effective potential.

Note that $V'_{eff}$ is zero at $\phi=0$ except when $\xi$ is such
that the pole in $V'_{eff}$ occurs at $\phi=0$.
$V'_{eff}$ has poles when $1/2-\sqrt{\Delta}=-1, -2, -3, \ldots$.
This translates into poles at (see Eq.(\ref{eq:Delta})),
$V''_{tree}/H^2=0,-4,-10,\ldots,-(n-1)(n+2),\ldots$.  Therefore a pole
will occur at $\phi=0$ if $\xi(H)=0,-1/3,\ldots,-(n-1)(n+2)/12,\ldots$.
For these values of $\xi(H)$, perturbation theory breaksdown
in the neighborhood of the origin.  Since in all cases where
perturbation theory is valid in the neighborhood of the origin,
$V'_{eff}(0)=0$, the origin is an extremum of $V_{eff}$.
Now if we can show that $V''_{eff}(0)<0$ for $\xi(H)<0$, then we will
have proven that a necessary condition for the symmetric case is that
$\xi(H)>0$.

In the neighbourhood of the origin, $V'_{eff}$ is
given by Eq.(\ref{eq:Vprime_H}).  Differentiating this expression with
respect to $\phi$ we obtain,
\begin{eqnarray}
V''_{eff}&=R\left(\xi-\f{\lambda}{384 \pi^2}
\left[\left(\Delta-\f{1}{4}\right)f-\f{1}{3}\right]\right) +
{}~~~~~~~~~~~~~~~~~~~~~~~~~~~~~~~~~~~~\nonumber \\
&~~\f{\lambda}{2}\phi^2\left(1+\f{\lambda}{16 \pi^2}
\left[f+\left(\Delta-\f{1}{4}\right)
\f{\Psi'\left(\f{1}{2}+\sqrt{\Delta}\right) -
\Psi'\left(\f{1}{2}-\sqrt{\Delta}\right)}{2\sqrt{\Delta}}\right]\right)
\end{eqnarray}
where $f$ represents the expression inside the curly brackets in
Eq.(\ref{eq:Vprime_H}), and all the couplings are normalized at
$\mu^2=H^2$.  This expression is regular at $\phi=0$ except for
the values of $\xi$ that we have listed above.  For $\xi$ not equal to
one of these values, we have,
\beq
V''_{eff}(0)=R\left(\xi+\f{\lambda}{384\pi^2}
\left\{12\left(\xi-\f{1}{6}\right)[f]_{\phi=0}+\f{1}{3}\right\}\right)
\label{eq:vpp0}
\eeq
where,
\beq
[f]_{\phi=0}=\Psi\left(\f{1}{2}+\sqrt{\f{9}{4}-12\xi}\right)+
\Psi\left(\f{1}{2}-\sqrt{\f{9}{4}-12\xi}\right)-1
\eeq
The plot of $V''_{eff}(0)$ versus $\xi(H)$ is shown for $\lambda(H)=1$
in Figure 2.
\begin{figure}
\centering
\epsfxsize=4in
\hspace*{0in}
\epsffile{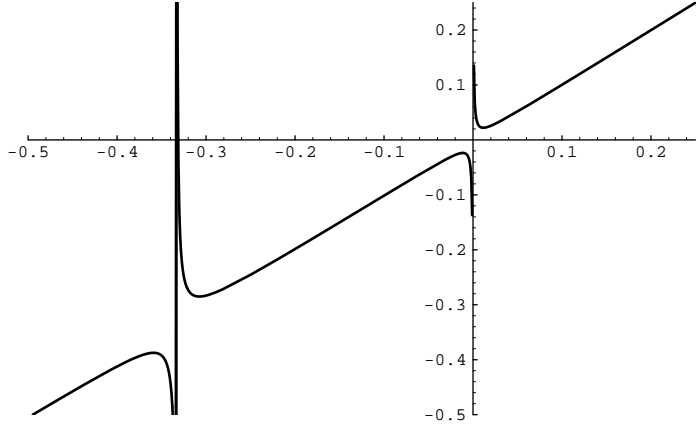}
\caption{Plot of $V''_{eff}(0)$ verses $\xi(H)$ for $\lambda(H)=1$.}
\label{fig2}
\end{figure}
It is evident that for $\xi(H)<0$,
$V''_{eff}(0)<0$, except for values of $\xi(H)$ when
perturbation theory is not valid at the origin.  Thus a
necessary condition for the symmetric case is that $\xi(H)>0$.

As we mentioned earlier, $\xi(H)$ can not be too small.
In order to get an estimate on how large $\xi(H)$ must
be for perturbation theory to be valid, we again look at
$V''_{eff}(0)$.  For perturbation theory to be valid, a sufficient
condition is (see Eq.(\ref{eq:vpp0})),
\beq
\left\vert\f{\lambda}{384\pi^2}
\left\{12\left(\xi-\f{1}{6}\right)[f]_{\phi=0}+\f{1}{3}\right\}
\right\vert \ll \xi
\eeq
For $\lambda$ of $O(1)$, this term will get large only near the pole.
Therefore we may approximate this expression with just the pole term.
We can extract the pole at $\xi=0$ from the function
$\Psi(1/2-\sqrt{9/4-12 \xi})$ using the following property of the
digamma function \cite{Ab&St},
\beq
\Psi(z+1)=\Psi(z)+\f{1}{z}
\eeq
Thus we obtain the condition,
\beq
\left\vert\f{\lambda}{384 \pi^2}12\left(-\f{1}{6}\right)
\left(-\f{1}{4\xi}\right)\right\vert \ll \xi
\eeq
giving,
\beq
\xi^2(H) \gg \f{\lambda(H)}{768 \pi^2}
\label{eq:condition_1}
\eeq
Therefore for all $\xi(H)>0$ that satisfies the above condition, we
have the symmetric case.

Now we will treat the spontaneously broken case.  At this point we
introduce the following notation for brevity.
$${\tilde \xi}(\phi) = \left\{\begin{array}{ll}
			    \xi(\phi) &{\rm if}~\phi\ge H \\
			    \xi(H)    &{\rm if}~\phi<H \end{array}
\right. $$
$${\tilde \lambda}(\phi) = \left\{\begin{array}{ll}
			    \lambda(\phi) &{\rm if}~\phi\ge H \\
			    \lambda(H)    &{\rm if}~\phi<H \end{array}
\right. $$
Now $V_0$ has minima away from the origin.  These minima occur at
$\phi=\pm v_0$ given by
\beq
v_0^2=\f{-6\xit(v_0)}{\lambdat(v_0)}R=\f{-72\xit(v_0)}{\lambdat(v_0)}H^2.
\label{eq:v_0}
\eeq
In the cases where perturbation theory can be trusted, the one-loop
corrections make only small shifts in the positions of these minima.
For some values of the couplings,
upon the inclusion of the full one-loop corrections, we
find that perturbation theory breaks down in the neighborhood of
$v_0$.  As we mentioned earlier, this is because the pole in
$V'_{eff}$ that occurs when the curvature of the tree potential
vanishes significantly affects the structure of $V_{eff}$ near this
apparent minimum.

To see this in greater detail, let us examine the effective potential
in the neighborhood of $\phi=+v_0$.
$V'_{eff}$ has poles at $\sqrt{\Deltat(\phi)}=3/2, 5/2, 7/2, \ldots$.
The pole that lies closest to the minimum is at
$\sqrt{\Deltat(\phi)}=3/2$, that is, when $V''_{tree}(\phi)=0$.
We extract the pole term as we did in the previous case.
The pole term is,
\beq
[V'_{eff}]_{\rm pole term}=\f{\lambdat(\phi) R \phi}{384 \pi^2}
\left(\Deltat(\phi) - \f{1}{4}\right)
\f{1}{\f{3}{2}-\sqrt{\Deltat(\phi)}}
\eeq
Now we want to find the range of $\phi$ over which the pole term is
dominant. That is, to a good approximation,
the range of $\phi$ over which,
\beq
\left|[V'_{eff}]_{\rm pole term}\right| \geq
\left|V'_0\right|
\eeq
where,
\beq
V'_0 \equiv
\phi\left[\xit(\phi)R+\f{\lambdat(\phi)}{6}\phi^2\right]
\eeq
To determine the limits of the allowed range of $\phi$,
we need to solve,
\beq
\left|[V'_{eff}]_{\rm pole term}\right| =
\left|V'_0\right|
\eeq
Approximating all the couplings with their values at the pole, and
approximating the L.H.S. with the first term of its Laurent expansion,
we obtain,
\beq
\left|\f{\lambdat(\phi_p)}{64 \pi^2}
\f{R\phi_p}{12\left(\xit(\phi_p)
+\f{\lambdat(\phi_p)}{2}\f{\phi^2}{R}\right)}\right| =
\left|\f{2}{3}\xit(\phi_p)R\phi_p\right|
\label{eq:equal}
\eeq
where $\phi_p$ is the position of the pole given by,
\beq
\phi_p^2=\f{-2\xit(\phi_p)R}{\lambdat(\phi_p)}
\eeq
The two solutions to (\ref{eq:equal}) are,
\beq
\phi_\pm^2=\phi_p^2 \pm \f{R}{256 \pi^2 |\xit(\phi_p)|}
\eeq
Thus for the dominance of the pole to die out well before
$\phi=v_1$, where $v_1$ is the minimum of the full one-loop effective
potential, we need,
\beq
\phi_+^2-\phi_p^2 \ll v_1^2-\phi_p^2
\eeq
That is,
\beq
\f{1}{256 \pi^2 |\xit(\phi_p)|} \ll
\f{v_1^2}{R}-\f{-2\xit(\phi_p)}{\lambdat(\phi_p)}
\eeq
If we make the approximations, $v_1^2\simeq v_0^2$,
$\xit(\phi_p)\simeq \xit(v_0)$, and $\lambdat(\phi_p)\simeq
\lambdat(v_0)$, the above condition becomes,
\beq
64 \xit^2(v_0) \gg \f{\lambdat(v_0)}{16 \pi^2}
\label{eq:condition_2}
\eeq
Note that if $v_0<H$, $\xit(\phi_p)=\xit(v_0)=\xi(H)$, and likewise for
$\lambdat$.  This condition is essentially the same as
(\ref{eq:condition_1}) for the case where $\phi_0=0$.
As long as (\ref{eq:condition_2}) is satisfied, the pole does not dominate
the form of the effective potential near the minimum, and we can trust
the one-loop effective potential near the minimum.

It is interesting to note that the ratio of $\xi^2$ and $\lambda$
appears in the expression for the value of the tree potential at its
minimum.  If we normalize the couplings at the scale of the minimum,
the value of the tree potential at $\phi=v_0$ relative to the value of
the potential at $\phi=0$ is,
\beq
V_{tree}(\phi=v_0)=-\f{3}{2}\f{\xit^2(v_0) R^2}{\lambdat(v_0)}.
\eeq
If the inequality (\ref{eq:condition_2}) is satisfied then,
\beq
V_{tree}(\phi=v_0) \ll -\f{3 R^2}{2048 \pi^2}.
\eeq
This condition implies that in the cases where perturbation theory is
valid, the value of the potential at the minimum is not degenerate or
``nearly'' degenerate with its value at the origin.

Thus we find that in all cases where perturbaton theory is valid,
$\vert\xit\vert$ (normalized at the appropriate scale as described
above) can not be too small.  What is too small is determined by
$\lambdat$ as given by Eq.(\ref{eq:condition_1}), and
Eq.(\ref{eq:condition_2}) for the two cases.   In any
case $\xit$ can not be zero at the minimum. Thus in
all cases where perturbation theory can be trusted,
the field will be non-minimally coupled at the minimum of
the effective potential.

\section{Conclusion}

We have analyzed the one-loop, renormalization-group-improved effective
potential for the $\lambda \phi^4$ theory in de~Sitter space and have
characterized the ranges of the parameters that lead to the different
forms of the effective potential.  The important result we have
obtained is that in all cases where perturbation theory can be
trusted,
the field is necessarily non-minimally coupled at the minimum of
the effective potential.  Consequently, normalizing the theory at the
scale where $\xi$ vanishes will lead to the breakdown of perturbation
theory.

Although we have investigated only the $\lambda \phi^4$ theory here,
there is one aspect of this theory that is generally true in all
theories.  It is that once
interactions are included, $\xi=0$ is never a fixed point
\cite{bos}.  Therefore, although one would have to do a
complete analysis of the other theories to determine their precise
behavior, the fact that $\xi=0$ is not a fixed point makes
the assumption of a minimally coupled scalar field suspect or, at least,
unnatural.  Thus we conclude that, in perturbation theory, there
are no long-distance problems associated with an interacting massless
scalar in de~Sitter space since the field is necessarily non-minimally
coupled.

It is worth reflecting on the fact that gravitons resemble to some degree
the massless, minimally-coupled scalar to the extent that the free field
equation for the transverse, traceless modes of the metric tensor is the
same, so that the graviton propagator suffers the same infrared problems.
However, the minimal coupling of these modes is a result of gauge
invariance (general coordinate invariance) and is not artificial. In that
case, we believe that a de~Sitter invariant vacuum does not strictly
exist, although the resolution of the infrared problem may not be so
serious.\footnote{See Ref.~\cite{dez} for a discussion of the modification
of the vacuum state that we have in mind.} There remain however other
possible mechanisms that feed back on the metric,\cite{Mottola,Woodard}
and it is not yet clear whether the necessary modifications of the
de~Sitter background in quantum gravity are relevant to the solution of
the cosmological constant problem.  The point is that, unfortunately, when
interactions are taken into account, the scalar case is not a good
prototype for quantum gravity.
\vskip.3in
\centerline {\large \bf Acknowledgements}
\vskip.1in
\noindent
We would like to thank A. Linde and S. Odintsov for their comments, which led
us to restate our results on the infrared limit in the symmetric case.
\vfill\eject


\end{document}